# Deep analytics of atomically-resolved images: manifest and latent features

Maxim Ziatdinov[1,2,*], Ondrej Dyck[1,2], Artem Maksov[1,2,4], Bethany M. Hudak[2,3], Andrew R. Lupini[2,3], Jiaming Song[3], Paul C. Snijders[3], Rama K. Vasudevan[1,2], Stephen Jesse[1,2], and Sergei V. Kalinin[1,2]

*E-mail: ziatdinovmax@gmail.com

[1]*Center for Nanophase Materials Sciences,* [2]*Institute for Functional Imaging of Materials,* [3]*Materials Science and Technology Division, Oak Ridge National Laboratory, Oak Ridge TN 37831*

[4]*Bredesen Center for Interdisciplinary Research, University of Tennessee, Knoxville, Tennessee 37996*

**Abstract**

Recent advances in scanning transmission electron and scanning tunneling microscopies allow researchers to measure materials structural and electronic properties, such as atomic displacements and charge density modulations, at an Angstrom scale in real space. At the same time, the ability to quickly acquire large, high-resolution datasets has created a challenge for rapid physics-based analysis of images that typically contain several hundreds to several thousand atomic units. Here we demonstrate a universal deep-learning based framework for locating and characterizing atomic species in the lattice, which can be applied to different types of atomically resolved measurements on different materials. Specifically, by inspecting and categorizing features in the output layer of a convolutional neural network, we are able to detect structural and electronic "anomalies" associated with the presence of point defects in a tungsten disulfide monolayer, non-uniformity of the charge density distribution around specific lattice sites on the surface of strongly correlated oxides, and transition between different structural states of buckybowl molecules. We further extended our method towards tracking, from one image frame to another, minute distortions in the geometric shape of individual Si dumbbells in a 3-dimensional Si sample, which are associated with a motion of lattice defects and impurities. Due the applicability of our framework to both scanning tunneling microscopy and scanning transmission electron microscopy measurements, it can provide a fast and straightforward way towards creating a unified database of defect-property relationships from experimental data for each material.

**Introduction**

Nano- and atomic scale imaging and spectroscopy have become the hallmark of modern times, allowing unprecedented access to materials structure and functionalities.[1-3] Electron beam techniques such as scanning transmission electron microscopy (STEM) now allow single atom imaging[4] and probing the local states of quasiparticles and their interactions in the solid-state systems.[5,6] Meanwhile, scanning probe techniques, such as atomic force and scanning tunneling microscopy (STM), enable a broad range of studies of materials surfaces, including imaging of magnetic[7] and ferroelectric[8] domains, single molecule recognition for biological applications,[9] as well as atomically resolved imaging of physical and chemical processes on metallic and semiconducting surfaces.[10-12]

The continuous development in these techniques and emergence of *in-operando* imaging gives rise to progressively increasing volumes of data, as exemplified by the development of dynamic STEM,[13,14] where one can observe, in real time, changes in the structure of a sample under the external perturbations, as well as e-beam atomic manipulation techniques.[15-17] These developments necessitate rapid image analysis, ideally immediately during acquisition.

This task becomes particularly relevant in the context of recently discovered controlled (manually) e-beam modification of materials, such as movement of single[18-20] and multiple[21] Si atoms over several lattice sites in graphene. Particularly, in order to achieve fast and automated, feedback-based atomic-level manipulation of materials structure, it is necessary to "teach" a microscope to localize and recognize various atomic-scale objects (atoms, atomic columns, defects) from one STEM (or STM) frame to another in real time, as well as to interpret the newly created (during a manipulation stage) and previously unseen atomic features 'on-the-fly' in a way similar to how human operators work. One of the most promising method for creating such an AI microscope is based on deep artificial neural networks that are capable of learning from the image in a way somewhat similar to humans.[22,23]

Deep artificial neural networks have witnessed unprecedented success, progressing from recognizing pictures of cats and dogs on the Internet to defeating a human champion in one of the most sophisticated human games in the world in just about three years.[22] Deep learning (DL) is now being actively used in various engineering applications, such as object detection and recognition[24] and unsupervised text translation,[25] to name just a few. More recently, deep learning models have been utilized in the domain of theoretical physics for the analysis of data collected at the Large Hadron Collider[26] and detection of phase transitions in lattice models.[27,28]

There are three critical aspects to consider before applying DL analytics to scientific imaging. The first aspect is a network's topology. For atomic imaging, for example, it is crucial to locate accurately all the atomic species in the image, hence, the model's output layer must provide a pixel-wise classification map of the same size as the input image. The second critical aspect is the creation of an appropriate training set. The currently available DL models pre-trained on objects from the macroscopic world cannot be used for analysis of atomic species and defects. When creating a new training set specifically for atomically-resolved imaging, one of the key issues is a trade-off between a model's ability to generalize to different types of atomic structures (e.g. lattices with different periodicities and symmetries) and its accuracy (e.g. how precise is the localization and identification of atomic species and defects). The question is: shall one use a different model for each lattice type or instead create a universal model that is able to analyze all types of lattices? Because the former approach will not work for complex materials in which several phases (and interface regions) may be present in a single image, it is preferable to have one model capable of recognizing multiple types of lattices (and associated lattice disorder), provided that a trade-off between generalization and accuracy is acceptable for physics-based interpretation of the obtained results. Finally, the third aspect is whether further processing of the DL model output can potentially yield more information. Here, one must be aware of several important differences with the applications of DL in a "real" macroscopic world, particularly, with those DL models used for segmentation tasks, such as fully convolutional networks.[29] Indeed, while a fully convolutional neural network can be an optimal tool for many (static) image segmentation tasks,[29,30] the interpretation of the results of such segmentation for the macroscopic world applications is rather straightforward and requires little to no further analysis (that is, simple visualization of the output is sufficient in the most cases). The situation, however, is entirely different, even for static images, in the world of atoms and molecules, where small changes in shapes and relative positions might imply emergence of certain physical phenomena (e.g. formation of charge and/or magnetic ordering). In this case, the outputs of the DL model can be refined by using theory-based constraints, for example, by applying a theory-informed Markov model to the output of the last convolutional layer. In addition, exploratory data analysis and anomaly detection methods can be applied to the DL model output to uncover crucial information about atomic defects in the system and their impact on lattice properties.

Here we have developed a deep learning based approach that allows finding atomic species, as well as lattice defects, in the atomically-resolved images from different materials without specifying the type of materials lattice. It works with both STEM and STM images, and is not sensitive to noise and other common types of experimental image distortion. We further extended our framework in such a way that it allows the user to perform exploratory data analysis, as well as anomaly detection and unsupervised and semi-supervised clustering, on the output of the DL model to uncover information linked to materials atomic structure and electronic properties.

**Results and Discussion.**

A schematic of our approach for analysis of experimental data using a single DL model for different materials and types of atomic lattice is shown in Fig. 1. It starts with collecting experimental data (e.g. from STM or STEM observations) and feeding it into the deep convolutional neural network. The network has an encoder-decoder type of architecture[31] such that an input image and the model output have the same dimensions (a crucial requirement for accurate localization of atomic species), and is trained using simulated images of atoms. As the model does not contain fully-connected ("dense") layers, it is referred to as a fully convolutional neural network (FCN).[29] We removed hard periodic constraints during simulation of training images, which allowed us to train a network that can find atoms in different types of lattices (see Supplemental Material). The typical output of FCN for atomically-resolved images represents a set of well-defined circular features on uniform background, where each feature corresponds to individual atom and/or atomic column. The key here is that variations in geometrical shapes of these features are not random but are linked to physical properties of the system (shape of atomic columns) as well as possible experimental anomalies such as abrupt changes in the tip apex shape in STM. Hence, a simple exploratory data analysis of the FCN output layer as well as more advanced data analytics techniques such as unsupervised classification/clustering and anomaly detection can yield a wealth of information on material's properties. Furthermore, in addition to performing a pixel based analysis of the FCN output, it is possible to construct physical descriptors of the system using parameters of the extracted atomic contours (area, circularity, orientation/angle, etc.) as well as other relevant parameters such as the nearest and next nearest neighbor distances between the centers of extracted contours and column intensity (summed over small box area around the detected center), and to use these descriptors as an input into the above mentioned data analytical models.

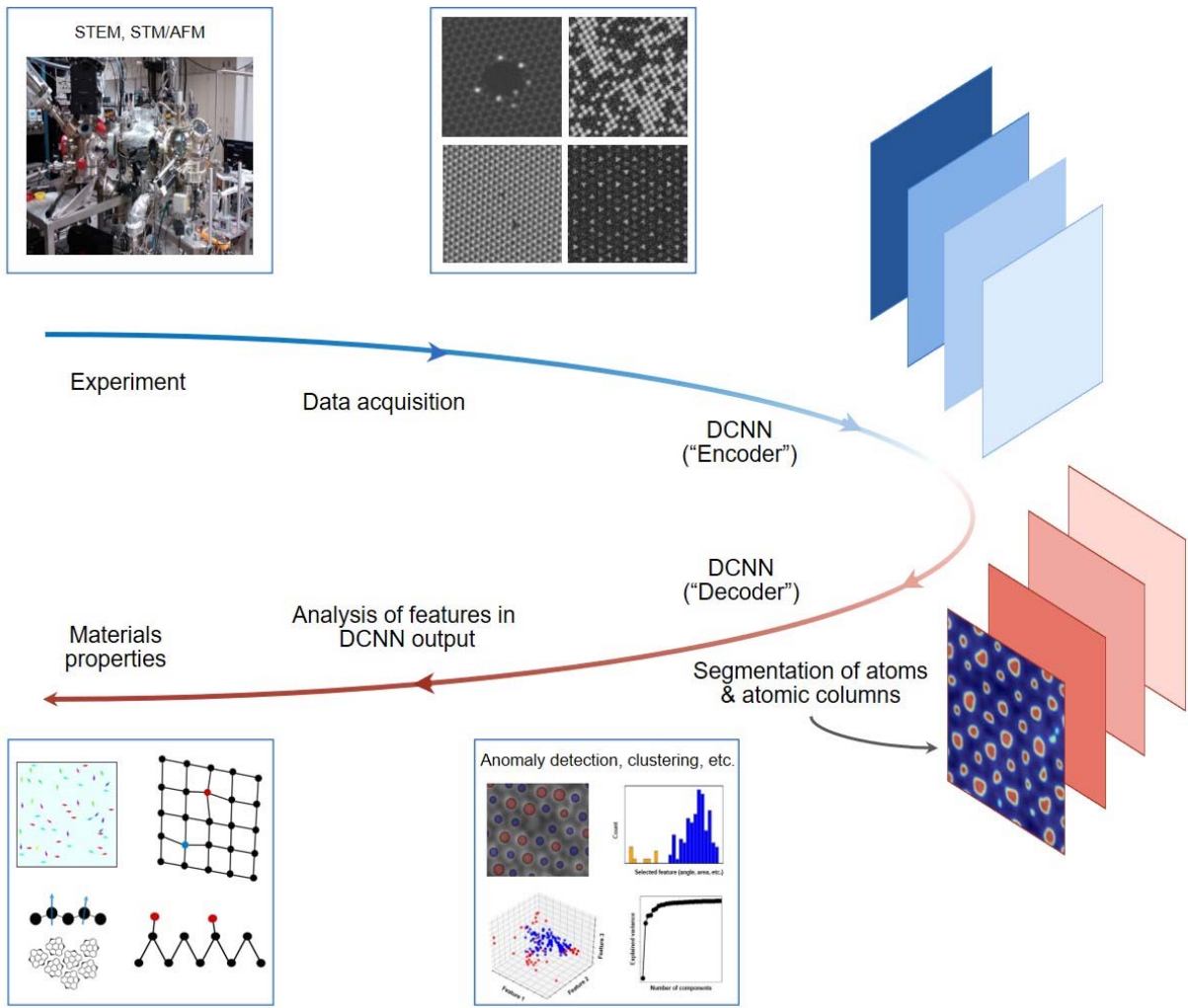

**Figure 1. Deep learning based identification of atoms, atomic columns and defects.** The experimental data is passed to a deep convolutional neural network (DCNN), which has an encoder-decoder type of structure (see Methods for details on network's architecture). The final (*softmax*) layer of the network typically shows a set of well-defined features which descriptors (intensity, area, circularity, etc.) are linked to physical processes defining shapes of atomic columns and/or certain experimental anomalies. By performing exploratory data analysis as well as applying appropriate clustering and/or anomaly detection algorithms to the *softmax* layer (model's output) one can learn important information about the atomic structure under consideration.

We start with illustrating how a deep learning model trained on the *simulated* data without hard periodic constraints can find atoms in different periodic lattices of real experimental images (Fig. 2). We first use our model to identify atomic positions in a STEM image obtained at an Al-Si interface region (Fig.

2a). Here the model was able to locate positions of atomic columns in both the Al lattice (bottom of the image) and in the Si lattice (top of the image), as we show in Fig. 2b. For comparison, if we use a model trained on a honeycomb, graphene-like lattice used in our earlier study,[32] it will find atomic positions in the Si lattice, but will fail to identify accurately atoms in Al lattice as it will try to "place" additional atoms in-between Al atoms to turn it into a honeycomb-like structure (see Supplementary Material).

To ensure the universality of our model for more different types of atomic lattices, as well as a different experimental technique, we applied it to an STM image of $La_{5/8}Ca_{3/8}MnO_3$ (LCMO) film, which has a relatively high degree of lattice disorder (Fig. 2c). Unlike the STEM image, which primarily reveals atomic positions, the STM image is associated with the electronic density of states around the Fermi level. While this can provide invaluable information about electronic behavior at the atomic scale, it also makes it more difficult for most image analysis techniques to locate atom-like or molecular features in the image and to categorize different structural and/or electronic states. In fact, automatic analysis of STM images is notoriously difficult compared to STEM, reflecting the more delocalized nature of STM signal.

In addition, on a somewhat deeper technical level, the STM images are usually characterized by large amounts of scars (strokes) due to interaction of an STM tip with a surface, which further complicates the task of finding atoms/molecules. Despite all these obstacles, our model was able to accurately identify positions of surface atoms in the un-processed image. We would like to emphasize that none of the three atomic lattice structures was explicitly included in our training set. This means that our model trained on simulated STEM data was able to learn the concept of what an atom/atomic column is and use it to find positions of atoms in various lattice structures in real experimental images, without being biased to one lattice type (and its periodicity). Furthermore, it was able to apply the learned concept of "atoms" to the analysis of an STM image, even though we did not train it specifically on STM data. This is important because to date the analytical methods developed for STEM have rarely been applied to STM, partially because the STM data is much "noisier" (due to tip-surface interaction), harder to match with simulated data (the image is a convolution of tip and sample density of states and is also affected strongly by out-of-plane surface deformations) and typically has lower pixel resolution. However, our results clearly show that the deep convolutional neural networks, which are not sensitive to pixel-level details and instead are able to capture "deeper" and more abstract concepts of the atomic lattices, can easily overcome these constraints.

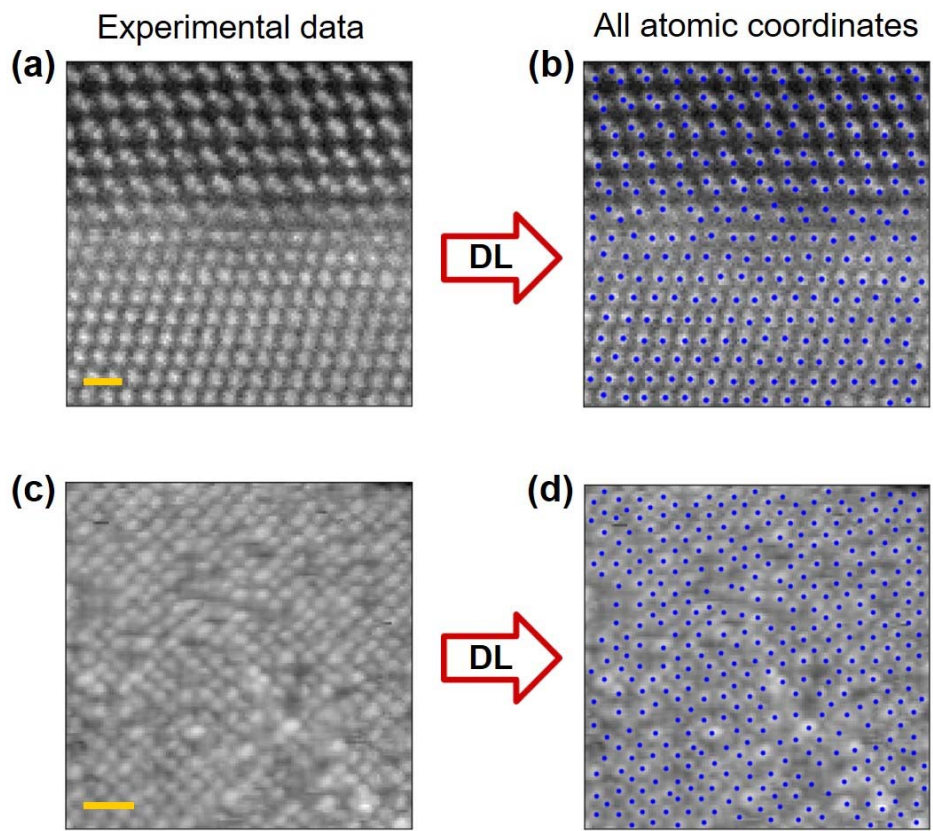

**FIGURE 2. Using a single DL model to identify positions of atoms/atomic columns in the images of different materials obtained from two different experimental techniques.** (a-b) STEM image of Al/Si interface (a) and identified positions of all atomic columns (b). (c-d) STM image of $La_{5/8}Ca_{3/8}MnO_3$ (LCMO) film (c) and identified positions of all surface atoms (d). Scale bar, 0.5 nm (a), 1.5 nm (c).

We next illustrate how a simple exploratory analysis of the DL model output can provide important information regarding types of atomic species in the image, using a STEM image of Mo-doped $WS_2$ (Fig. 3a) as an example. Note that the S atoms are hard to detect above the background noise in the STEM experiments on $WS_2$ and hence the lattice appears as having a trigonal structure instead of a honeycomb-like structure. While finding the likely locations of S atomic columns using FCN is possible in principle, we will not perform such an analysis in this study. The output of FCN (*softmax* layer) for this image is shown in Fig. 3b. It essentially represents a probability map of each pixel belonging to an atom/atomic column. The key observation here is that variation in shapes of circular-like features in the *softmax* output is not random, but is linked to actual changes in the shape/intensity of atomic projections in the STEM image. As such, features representing dopants/contaminations with lower (higher) Z are characterized by a smaller (larger) area (Fig. 3b). Note that the model was *not* specifically trained to return features of different sizes for atoms with different intensities. The analysis of other relevant geometric parameters such as circularity may provide additional information about lattice distortions, "switching" during the scan (particularly in STM images), or even about some more subtle effects such as a character of rapid atom oscillations in the shallow potential well of a 2D system.

We now illustrate how the search for "anomalies" in the output of DL model can be in principle automated. Once we establish a probability of each pixel belonging to an atom or to a background (that is, being relevant or irrelevant), we can extract a contour of each identified feature associated with atoms and/or atomic columns and characterize it based on its geometric descriptors such as area, circularity, orientation/angle, etc. For STEM image of the doped $WS_2$, a plot of area and circularity descriptors for all the extracted contours as well as associated histograms are shown in Fig. 3c.

We can then classify features from DL output based on the constructed descriptors using clustering techniques and/or anomaly detection methods. For data in Fig. 3a, the anomalies extracted *via* one class support vector machine method are plotted along with the regular lattice in Fig. 3d. The red circles identify what is likely to be Mo dopants (lower $Z$ atomic number compared to W), whereas the green circles correspond to "contaminations"/adatoms that frequently form on $WS_2$ surface under e-beam irradiation. In Fig. 4 we illustrated several more representative examples of successful application of this approach to STEM experimental data. Specifically, in Fig. 4a-b we show another STEM images of Mo-doped $WS_2$ for which, in addition to finding all dopants and atomic-size contaminations, our model is able to describe accurately a non-trivial shape of edges (*e.g.*, presence of both regular lattice atoms and dopants at the edge in Fig. 4a,d or considerable irregularities in the edge shape in Fig. 4b,e)**.** In Fig. 4c,f we are able to identify accurately positions of Bi dopants in a Si lattice, as well as locations of Si atomic species themselves.

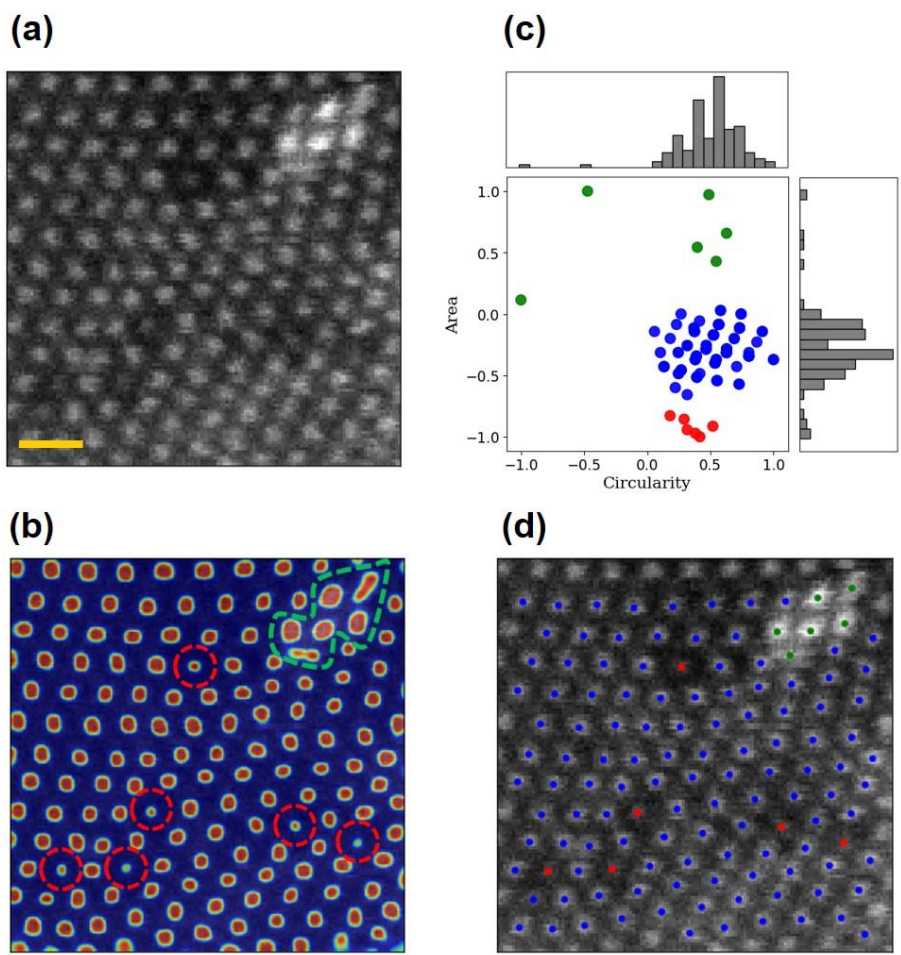

**FIGURE 3. Identification of anomalies in softmax layer of DL model.** (a) Experimental STEM image of Mo-doped $WS_2$. (b) *Softmax* layer of DL model for image in (a). "Contaminations"/adatoms and Mo dopant are shown by green and red dotted lines, respectively. (c) Plot of descriptors (normalized) associated with area and circularity of individual features in the *softmax* layer with corresponding histograms for each descriptor. The anomaly detection method (one class support vector machine) followed by semi-manual refinement (1 data point removed from anomaly "cluster") allowed to separate the extracted descriptors into clusters associated with "contaminations"/adatoms (green) and Mo dopants (red). (d) Atomic positions overlaid on the experimental image (coloring scheme the same as in (c)). Scale bar, 0.5 nm.

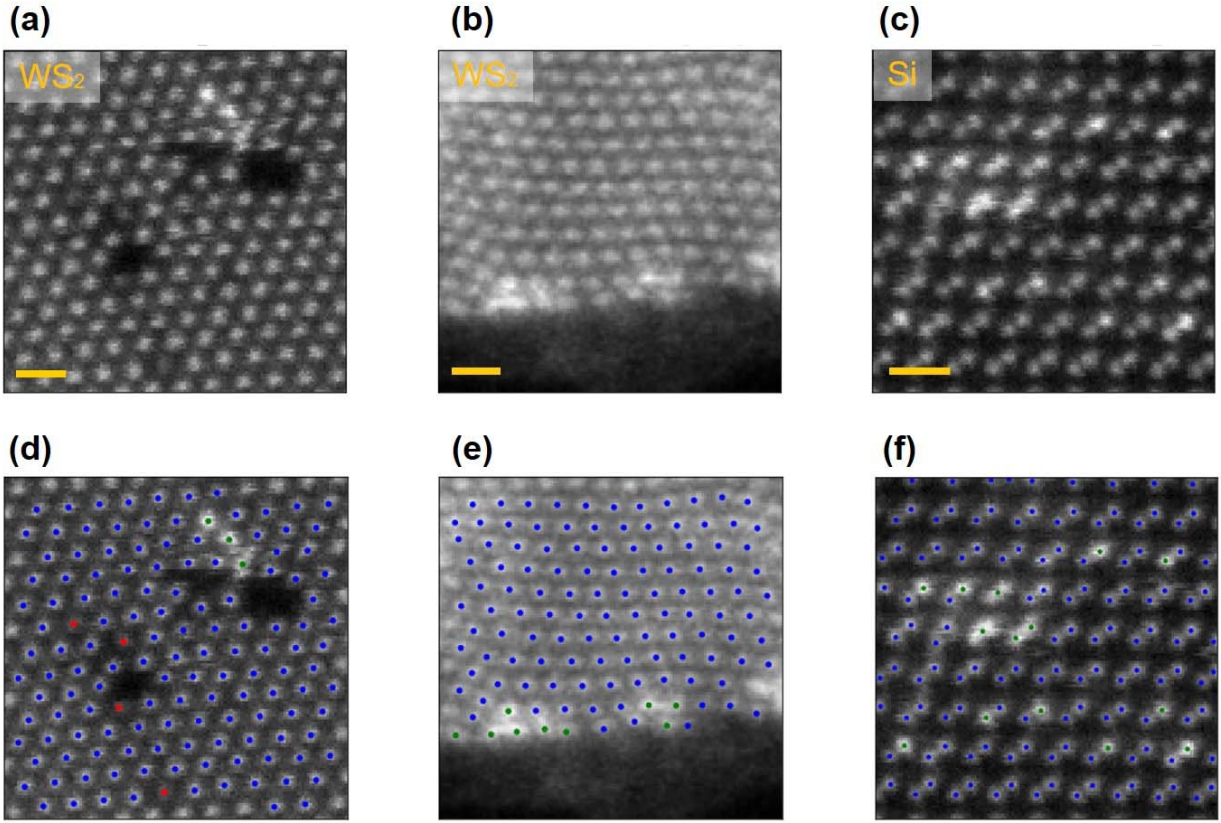

**FIGURE 4. Identification of lattice atoms and lattice disorder from STEM images using a single DL model.** (a-c) Experimental STEM images of $WS_2$ with small holes in the lattice (a) and step edge (b), and Si with Bi dopants (c). (d-f) Corresponding atomic positions (colormap meaning the same as in Fig. 3). All scale bars are 0.5 nm.

In STEM experiments on 3-dimensional materials, variations in the shapes or orientations of atomic columns may originate from atomic re-arrangements along the columns that are too small to be resolved as separated features but which might represent important structural changes. For example, in oxide materials, column shapes relate to the rotation of structural units.[33] We now illustrate how our DL-based model can be used to extract relevant information about the geometrical parameters of such distortions. As a specific test case, we use STEM images of a Si lattice with Bi dopants (Fig. 5). Here, in addition to non-uniformity introduced to the Si lattice by Bi dopants, which introduce significant local strain, we expect additional forms of lattice "disorder" caused by atomic motion and lattice reconstructions under the e-beam impact.[34] In this example, the sample reconstructs and changes during the experiment, making tracking of dopants and understanding the material's response a key problem.

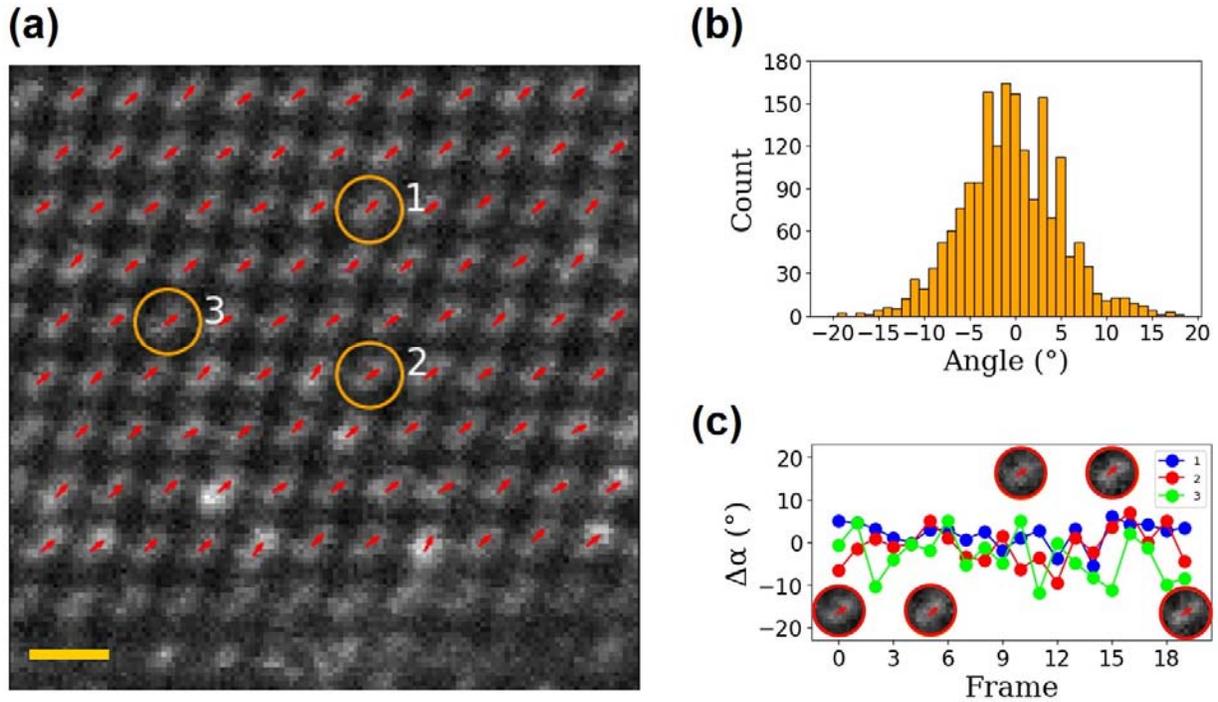

**FIGURE 5. Analysis of atomic columns distortion in 3-dimensional Si sample by tracking orientation of individual Si dumbbells.** (a) First (out of twenty) frame of STEM movie on Bi-doped Si sample. Orientation of Si dumbbells extracted from the output of DL model is shown by red arrows. (b) Distribution of Si dumbbell angles (with respect to the [001] direction) for twenty movie frames. (c) Tracking changes in orientation of individual Si dumbbells (orange circles in (a)) across twenty movie frames (count starts from zero). For several selected frames, the orientation of dumbbell #2 is shown by the red arrow overlaid on top of experimental data. Scale bar, 0.5 nm.

By tuning the threshold intensity value used for the extraction of atomic contours from the DL model output we can make a transition from analysis of individual Si columns to analysis of individual Si dumbbells (see Supplementary Material). Specifically, we approximate each extracted Si dumbbell contour with an ellipse and analyze its orientation (angle) with respect to the chosen frame of reference (Fig. 5a). As our data is now represented by the collection of well-defined contours on an "empty" background, we can easily track changes in the geometrical descriptors of a specified contour (or all contours at once) through multiple STEM frames (the STEM "movie") as we demonstrated in Fig. 5c. To track each "atomic contour" during the movie analysis (or acquisition) we first store the coordinates of the centers of all contours from the first frame and then locate the same contours in the subsequent frames by finding the closest contour for each pair of coordinates. This approach works if an overall shift between the first and the last frame due to sample drift is smaller than a half-distance between two neighboring dumbbells. For larger sample drifts one must update the stored coordinates after successfully locating the same contour(s) in each frame. This same idea can be used "in reverse" for drift correction between consecutive frames: in cases where there is either significant structural rearrangement or a repeating lattice, accurate alignment of sequential frames remains a difficult problem. The ability to obtain detailed information on the changes in orientation of individual structural units in an automated fashion can provide important clues about the types of disorder present in the material's volume. It may be particularly useful for analysis of long-range correlations in complex 3D systems with various types of charge and/or magnetic (dis-)order, such as complex cobaltates.[35] Furthermore, as we show below, this approach can be easily extended towards analysis of non-uniform charge density distribution in STM images of strongly correlated systems.

We proceed to applying the same DL model to analysis of STM images. Specifically, we apply our DL model to the STM data on an LCMO film. This material is characterized by a complex phase diagram with competing structural, chemical and electronic/magnetic orders which to date remains inadequately understood.[36] In some cases, such a competition can lead to formation of full or partial charge order characterized by a non-trivial shape and distribution of charge density "blobs".[37] Even without formation of charge order, analysis of charge density distribution around different lattice positions can provide important clues on local chemical environment and electronic environment. For example, since the STM images are imaging O atoms (these have the highest density of states near the Fermi level[38]), one can expect that depending on the local cation environment, the imaged atom may display elongated features caused by altered local electronic density, which would be expected when comparing $La^{3+}$ to $Ca^{2+}$ in this system. The raw experimental STM image is shown in Fig. 6a. In Fig. 6b we plotted the anomalies in the distribution of charge density at atomic lattice sites, detected by a one class support vector machine and associated with variation in circularity and area of individual features. The detected anomalies are plotted together with the experimental image and "atomic contours" in Fig 6c and 6d, respectively. Interestingly, by more closely analyzing the larger features (marked green) we found that the charge density appears to be "leaking" in one direction (that is, charge distribution is not identical at the ends of blob's long axis) which can be associated with non-homogeneous local chemical environment.

We finally show that it is possible to use our model for analyzing molecule resolved STM data. Specifically, in Fig. 7 we demonstrate a DL based analysis of STM image of sumanene molecules (also called buckybowls) on a gold substrate. The molecules can have different orientation with respect to the substrate (azimuthal rotations) and, under certain experimental conditions, can switch between their bowl-up and bowl-down structural conformations.[39] Interestingly, our model, which was trained to find atoms, was able to locate local maxima (hereafter, "blobs") in charge density distribution inside molecules (Fig. 7b). The latter is a product of hybridization of the molecular orbitals with substrate states and when viewed through the STM can be also affected by electronic and structural state of an STM tip apex, as well as by the out-of-plane deformations/curvature. The ability of our model to see these characteristic "blobs" can be viewed as if model treats each molecule as a cluster of "atoms" (not to be confused with actual, much larger number atoms inside a molecule). For the unperturbed molecular structure, the positions of such "blobs" within each molecule allow us to determine an orientation (azimuthal rotation) of the molecule with respect to the substrate (blue markers in Fig. 7b-e). On the other hand, the arrangement of "blobs" within each molecule is altered for a molecule that undergoes a transition between different structural (and/or rotational) states. We found that such a transition can be detected by our model (red markers in Fig. 7b and 7c). This ability to detect a transition between different molecular states via a deep learning framework may prove to be an invaluable tool for single molecule manipulations in future fully autonomous microscopes.

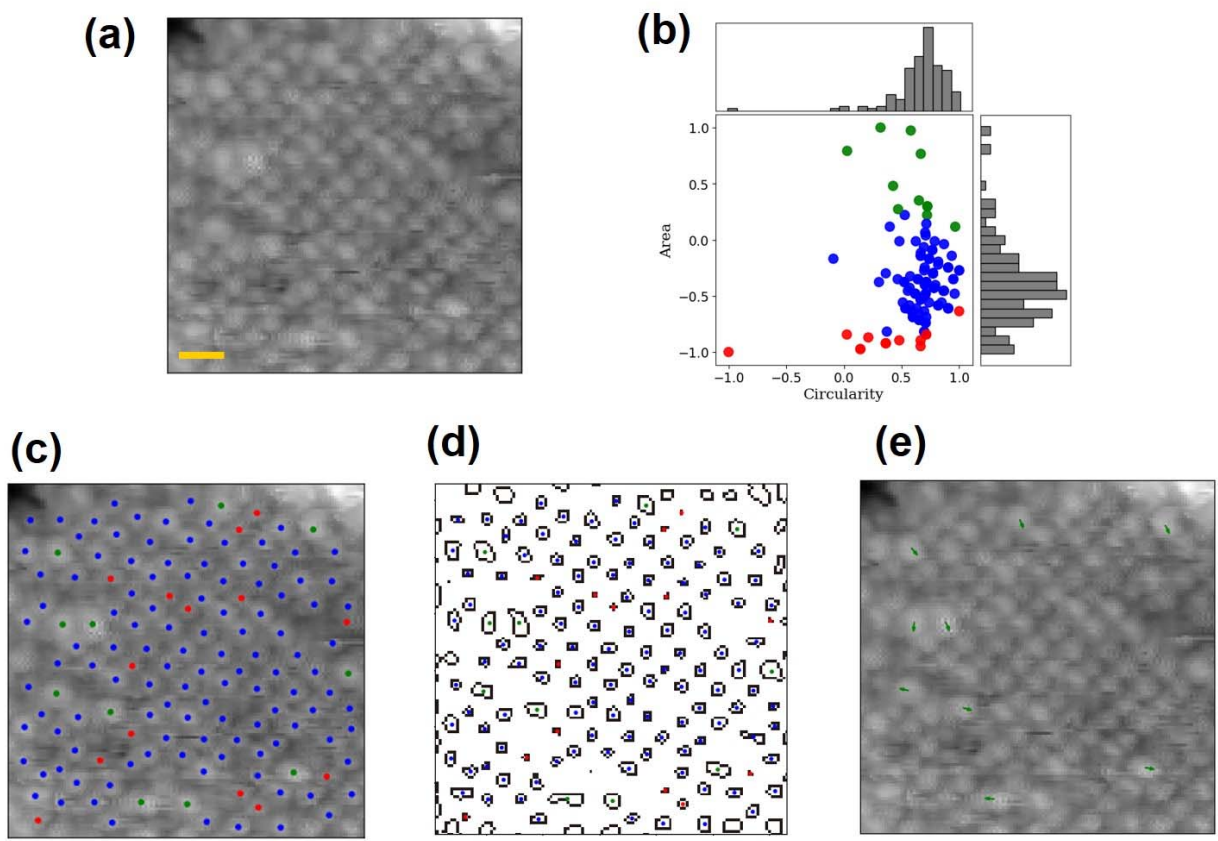

**FIGURE 6. Identification of anomalies in the output of DL model for STM data.** (a) Experimental STM image of LCMO surface. (b) Plot of geometrical descriptors (normalized) associated with area and circularity of individual features in the *softmax* layer with a corresponding histogram for each descriptor. (c) Atomic positions overlaid on the experimental image. The color labels are based on data analysis

shown in (b). (d) Visualization of "atomic contours", which in this case provides information about non-uniform distribution of charge on the surface. (e) Green arrows show schematically the distortion of electronic charge density for atoms labeled green in (c) and (d). Scale bar, 1 nm.

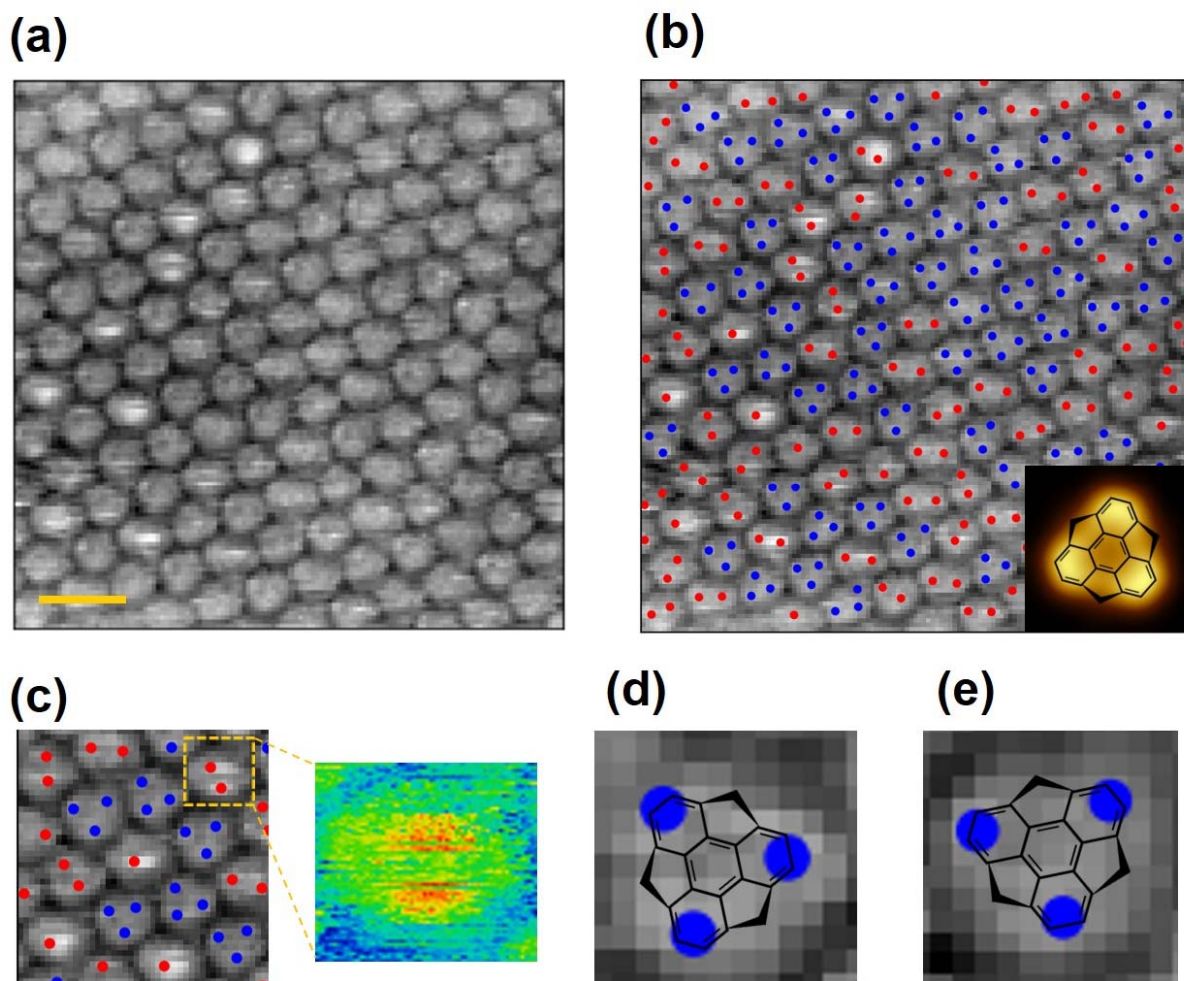

**FIGURE 7. Identifying molecular rotations and detecting a switching between different molecular conformations via DL.** (a) Experimental STM images of sumanene monolayer on gold surface. (b) Decoded molecular configurations. Molecules that do (do not) undergo a switching between different conformations are marked red (blue). The positions of blue markers correspond to a molecular orientation on the substrate as can be clearly seen from a zoomed area in (c). Zoomed-in illustrative example of a molecule undergoing a switching event during the scan is shown for one of molecules in (c). (d-e) Molecular chemical structure overlaid on top of the decoded structure for two different orientations. The overlay was done according to the distribution of charge density maxima determined from DFT STM simulations of molecules on the gold substrate[40] (see inset in (b)). Note that a match is not perfect in part due to the model's uncertainty, but also because of interaction

between molecules and STM tip as well as potential admixture of different rotational states and the associated dynamic averaging in STM experiment. Scale bar, 2 nm.

**Conclusions and Outlook**

In summary, we have developed a universal deep learning based analytical framework for fast and accurate analysis of atomic lattice structures across a wide range of materials. Within this framework, we first utilize a deep convolutional neural network for locating and extracting contours of individual atoms and atomic columns from the experimental images. We then construct and analyze geometrical descriptors of the extracted contours such as area, circularity, and orientation (angle) via exploratory data analysis and unsupervised anomaly detection/clustering. The variations in the contour's shape are linked to materials structural and electronic properties, such as presence and motion of dopants and contaminations on the surface and in the bulk, as well as a non-uniform spatial distribution of electronic charge density in the vicinity of certain surface lattice atoms. We further extended our approach towards analyzing rotations and detecting conformational changes of molecules and tracking specific geometrical descriptors associated with the selected atomic columns in STEM "movies" of 3D material. We foresee that due to its ability to quickly analyze raw data obtained from different types of atomically-resolved experimental measurements (structural data from STEM, electronic properties from STM), the created framework opens a pathway towards creation of comprehensive libraries of defect-property relationships for each technologically relevant nanoscale material, with the ultimate goal of making these open and discoverable across condensed matter physics and materials science communities. Such libraries will be also used to train and test state-of-the-art machine learning techniques for the materials science. In addition, they can be complemented by already available in literature experimental data, for example, by searching ("mining") the Internet for specific atomically-resolved images and then analyzing them via the same model as used for the experimental data obtained directly from a microscope.

Finally, we believe that our work represents an important step towards creation of AI framework capable of understanding and modifying the atomic and molecular structures. In the future research, we plan to build upon the DL atom/defect finder to include automatic lattice symmetry and defect type identification, which will allow extracting a complete information about materials structural and electronic parameters from microscopy images in real time. This will provide the final part of the AI framework, capable of modifying materials to match the desired structure, thus enabling the ultimate goal of nanotechnology – automated atomic scale manufacturing.

**Methods:**

**STEM experiment**

STEM imaging was performed using a Nion UltraSTEM U100 STEM operated at 60 kV and 100kV. A Nion UltraSTEM U200 aberration-corrected scanning transmission electron microscope, operating at 200 kV was used for the images of Bi-doped Si. The images were acquired in high angle annular dark field (HAADF) imaging mode and were introduced to the DL network without any post processing.

**STM experiment**

STM imaging of LCMO films was performed *in situ* at room temperature with an Omicron VT system at an operating pressure $< 3\times10^{-10}$ Torr using mechanically cut Pt/Ir tips in constant current mode. The images were taken with the tip negatively biased in respect to a grounded sample. STM imaging of sumanene molecules was performed at room temperature in a constant-current mode using the Japan Electron Optics Laboratory (JEOL) JSPM-4500S system at an operating pressure $\sim 10^{-11}$ Torr. The STM tips were prepared by electrochemical etching of tungsten wire and were further cleaned by $Ar^+$ ion sputtering (ion energy = 1.0–3.5 keV).

**Deep Learning**

The DL networks were implemented using Keras 2.0 (https://keras.io) Python deep learning library, with the TensorFlow backend. The convolutional neural network had an encoder-decoder type of structure. The first block of encoder consisted of two back to back convolutional layers (20 kernels of size 3 × 3 and stride of 1 in each layer) followed by max-pooling layer (size 2 × 2, stride 2), while the second block had three back to back convolutional layers (number of kernels in each layer 40, size 3 × 3, stride 1) followed by max-pooling layer (size 2 × 2, stride 2). The decoder part had the same structure but in the reversed order and was followed by a dropout layer (rate set to 0.25). The last two convolutional layers had the number of kernels equal to the number of classes to be determined (size 3×3 (pre-last layer) and 1×1 (last layer) and stride 1). The feature maps from the final convolutional layer of the network were fed into a *softmax* classifier for pixel-wise classification, providing us with information on the probability of each pixel belonging to an atom. The Adam optimizer[41] was used with categorical cross-entropy as the loss function. The initial simulated STEM images produced by a Multislice algorithm[42] were augmented to generate a large enough training set of about 3000 images. The augmentation was performed by applying random horizontal/vertical shifts, rotations, zooming-in/out and shear transformations. Each image was also corrupted with Poisson noise and blurring in the ranges comparable to those typically observed in experiments. The accuracy of our model on the test set was about 97 %. The extraction of contours and associated geometrical descriptors was performed in part using the Open Source Computer Vision Library (https://opencv.org).

**Supporting Information**

Example of training images, model accuracy plots, additional illustrations/discussion on how neural networks recognize various atomic features in STM and STEM data, as well as IPython notebooks with code for constructing and training a neural network model, extracting coordinates of all atoms, constructing geometrical descriptors from the output of the DL model and tracking how they change for the selected "atomic contours" in STEM movies.

**Acknowledgements:**

The authors thank Dr. Leonard from Oak Ridge National Laboratory for providing the STEM image of Al/Si interface. MZ acknowledges Prof. Sakurai from Osaka University for synthesis of the sumanene molecules, and Dr. Fujii and Prof. Kiguchi from Tokyo Institute of Technology for their assistance in STM measurements. MZ and AM thank Dr. Laanait from Oak Ridge National Laboratory for insightful discussions regarding applications of deep learning in atomically resolved imaging. This research was sponsored by the Division of Materials Sciences and Engineering, Office of Science, Basic Energy Sciences, US Department of Energy (MZ, RVK, ARL and SVK). Research was conducted at the Center for Nanophase Materials Sciences, which is a DOE Office of Science User Facility, and was also supported by the Laboratory Directed Research and Development Program of Oak Ridge National Laboratory, managed by UT-Battelle, LLC, for the U.S. Department of Energy (OD, BMH, PCS, JS, SJ).

**References:**

(1) Pennycook, S. J.; Nellist, P. D.: *Scanning Transmission Electron Microscopy*; Springer-Verlag New York, 2011.

(2) Moore, A. M.; Weiss, P. S. Functional and Spectroscopic Measurements with Scanning Tunneling Microscopy. *Annual Review of Analytical Chemistry* **2008**, *1*, 857-882.

(3) Zandvliet, H. J. W.; van Houselt, A. Scanning Tunneling Spectroscopy. *Annual Review of Analytical Chemistry* **2009**, *2*, 37-55.

(4)	Ishikawa, R.; Pennycook, S. J.; Lupini, A. R.; Findlay, S. D.; Shibata, N.; Ikuhara, Y. Single atom visibility in STEM optical depth sectioning. *Applied Physics Letters* **2016**, *109*, 163102.

(5)	Wei, J.; Jiang, N.; Xu, J.; Bai, X.; Liu, J. Strong Coupling between ZnO Excitons and Localized Surface Plasmons of Silver Nanoparticles Studied by STEM-EELS. *Nano Letters* **2015**, *15*, 5926-5931.

(6)	Tizei, L. H. G.; Lin, Y.-C.; Mukai, M.; Sawada, H.; Lu, A.-Y.; Li, L.-J.; Kimoto, K.; Suenaga, K. Exciton Mapping at Subwavelength Scales in Two-Dimensional Materials. *Physical Review Letters* **2015**, *114*, 107601.

(7)	Collins, L.; Belianinov, A.; Proksch, R.; Zuo, T.; Zhang, Y.; Liaw, P. K.; Kalinin, S. V.; Jesse, S. G-mode magnetic force microscopy: Separating magnetic and electrostatic interactions using big data analytics. *Applied Physics Letters* **2016**, *108*, 193103.

(8)	Kholkin, A. L.; Kalinin, S. V.; Roelofs, A.; Gruverman, A.: Review of Ferroelectric Domain Imaging by Piezoresponse Force Microscopy. In *Scanning Probe Microscopy: Electrical and Electromechanical Phenomena at the Nanoscale*; Kalinin, S., Gruverman, A., Eds.; Springer New York: New York, NY, 2007; pp 173-214.

(9)	Stroh, C.; Wang, H.; Bash, R.; Ashcroft, B.; Nelson, J.; Gruber, H.; Lohr, D.; Lindsay, S. M.; Hinterdorfer, P. Single-molecule recognition imaging microscopy. *Proceedings of the National Academy of Sciences of the United States of America* **2004**, *101*, 12503-12507.

(10)	Wang, Y.; Wong, D.; Shytov, A. V.; Brar, V. W.; Choi, S.; Wu, Q.; Tsai, H.-Z.; Regan, W.; Zettl, A.; Kawakami, R. K.; Louie, S. G.; Levitov, L. S.; Crommie, M. F. Observing Atomic Collapse Resonances in Artificial Nuclei on Graphene. *Science* **2013**, *340*, 734.

(11)	Kazuma, E.; Jung, J.; Ueba, H.; Trenary, M.; Kim, Y. Direct Pathway to Molecular Photodissociation on Metal Surfaces Using Visible Light. *Journal of the American Chemical Society* **2017**, *139*, 3115-3121.

(12)	Ziatdinov, M.; Banerjee, A.; Maksov, A.; Berlijn, T.; Zhou, W.; Cao, H. B.; Yan, J. Q.; Bridges, C. A.; Mandrus, D. G.; Nagler, S. E.; Baddorf, A. P.; Kalinin, S. V. Atomic-scale observation of structural and electronic orders in the layered compound α-RuCl3. *Nature Communications* **2016**, *7*, 13774.

(13)	Li, Y.; Zakharov, D.; Zhao, S.; Tappero, R.; Jung, U.; Elsen, A.; Baumann, P.; Nuzzo, R. G.; Stach, E. A.; Frenkel, A. I. Complex structural dynamics of nanocatalysts revealed in Operando conditions by correlated imaging and spectroscopy probes. *Nature Communications* **2015**, *6*, 7583.

(14)	Chi, M.; Wang, C.; Lei, Y.; Wang, G.; Li, D.; More, K. L.; Lupini, A.; Allard, L. F.; Markovic, N. M.; Stamenkovic, V. R. Surface faceting and elemental diffusion behaviour at atomic scale for alloy nanoparticles during in situ annealing. *Nature Communications* **2015**, *6*, 8925.

(15) Jesse, S.; He, Q.; Lupini, A. R.; Leonard, D. N.; Oxley, M. P.; Ovchinnikov, O.; Unocic, R. R.; Tselev, A.; Fuentes-Cabrera, M.; Sumpter, B. G.; Pennycook, S. J.; Kalinin, S. V.; Borisevich, A. Y. Atomic-Level Sculpting of Crystalline Oxides: Toward Bulk Nanofabrication with Single Atomic Plane Precision. *Small* **2015**, *11*, 5895-5900.

(16) Zhang, W. M.; Wang, Y. G.; Li, J.; Xue, J. M.; Ji, H.; Ouyang, Q.; Xu, J.; Zhang, Y. Controllable shrinking and shaping of silicon nitride nanopores under electron irradiation. *Applied Physics Letters* **2007**, *90*, 163102.

(17) Kalinin, S. V., Borisevich, A., Jesse, S. Fire up the atom forge. *Nature* **2016**, *539*, 485-487.

(18) Dyck, O.; Kim, S.; Kalinin, S. V.; Jesse, S. Single atom manipulation and control in a scanning transmission electron microscope. *arXiv preprint arXiv:1708.01523* **2017**.

(19) Susi, T.; Meyer, J. C.; Kotakoski, J. Manipulating low-dimensional materials down to the level of single atoms with electron irradiation. *Ultramicroscopy* **2017**, *180*, 163-172.

(20) Susi, T.; Tripathi, M.; Meyer, J. C.; Kotakoski, J. Electron-beam manipulation of Si dopants in graphene. *arXiv preprint arXiv:1712.08755* **2017**.

(21) Dyck, O.; Kim, S.; Jimenez-Izal, E.; Alexandrova, A. N.; Kalinin, S. V.; Jesse, S. Assembling Di-and Multiatomic Si Clusters in Graphene via Electron Beam Manipulation. *arXiv preprint arXiv:1710.09416* **2017**.

(22) LeCun, Y.; Bengio, Y.; Hinton, G. Deep learning. *Nature* **2015**, *521*, 436-444.

(23) Krizhevsky, A.; Sutskever, I.; Hinton, G.: ImageNet classification with deep convolutional neural networks. In *Proc. Advances in Neural Information Processing Systems*, 2012; Vol. 25; pp 1090-1098.

(24) Zhou, B.; Khosla, A.; Lapedriza, A.; Oliva, A.; Torralba, A.: Learning deep features for discriminative localization. In *Proceedings of the IEEE Conference on Computer Vision and Pattern Recognition*, 2016; pp 2921-2929.

(25) Artetxe, M.; Labaka, G.; Agirre, E.; Cho, K. Unsupervised Neural Machine Translation. *arXiv preprint arXiv:1710.11041* **2017**.

(26) Baldi, P.; Sadowski, P.; Whiteson, D. Searching for exotic particles in high-energy physics with deep learning. *Nature Communications* **2014**, *5*, 4308.

(27) van Nieuwenburg, E. P. L.; Liu, Y.-H.; Huber, S. D. Learning phase transitions by confusion. *Nat Phys* **2017**, *13*, 435-439.

(28) Carrasquilla, J.; Melko, R. G. Machine learning phases of matter. *Nat Phys* **2017**, *13*, 431-434.

(29) Long, J.; Shelhamer, E.; Darrell, T.: Fully convolutional networks for semantic segmentation. In *Computer Vision and Pattern Recognition (CVPR)*; IEEE: Boston, MA, USA, 2015.

(30) Badrinarayanan, V.; Handa, A.; Cipolla, R. SegNet: A Deep Convolutional Encoder-Decoder Architecture for Robust Semantic Pixel-Wise Labelling. *arXiv preprint arXiv:1505.07293* **2015**.

(31) Badrinarayanan, V.; Kendall, A.; Cipolla, R.: SegNet: A Deep Convolutional Encoder-Decoder Architecture for Scene Segmentation. In *IEEE Transactions on Pattern Analysis and Machine Intelligence*; IEEE, 2017.

(32) Ziatdinov, M.; Dyck, O.; Maksov, A.; Li, X.; Sang, X.; Xiao, K.; Unocic, R. R.; Vasudevan, R.; Jesse, S.; Kalinin, S. V. Deep Learning of Atomically Resolved Scanning Transmission Electron Microscopy Images: Chemical Identification and Tracking Local Transformations. *ACS Nano* **2017,** *11*, 12742–12752.

(33) He, Q.; Ishikawa, R.; Lupini, A. R.; Qiao, L.; Moon, E. J.; Ovchinnikov, O.; May, S. J.; Biegalski, M.; Borisevich, A. Y. Towards 3D mapping of BO6 Octahedron rotations at perovskite heterointerfaces, unit cell by unit cell. *ACS Nano* **2015,** *9*, 8412–8419.

(34) Ishikawa, R.; Mishra, R.; Lupini, A. R.; Findlay, S. D.; Taniguchi, T.; Pantelides, S. T.; Pennycook, S. J. Direct Observation of Dopant Atom Diffusion in a Bulk Semiconductor Crystal Enhanced by a Large Size Mismatch. *Physical Review Letters*, **2014**, *113*, 155501.

(35) Kishida, T.; Kapetanakis, M. D.; Yan, J.; Sales, B. C.; Pantelides, S. T.; Pennycook, S. J.; Chisholm, M. F. Magnetic Ordering in Sr3YCo4O10+x. *Scientific Reports* **2016**, *6*, 19762.

(36) Tokura, Y. Critical features of colossal magnetoresistive manganites. *Reports on Progress in Physics* **2006**, *69*, 797.

(37) Renner, C.; Aeppli, G.; Kim, B. G.; Soh, Y.-A.; Cheong, S. W. Atomic-scale images of charge ordering in a mixed-valence manganite. *Nature* **2002**, *416*, 518.

(38) Tselev, A.; Vasudevan, R. K.; Gianfrancesco A. G.; Qiao, L.; Ganesh, P.; Meyer T. L.; Lee, H. N.; Biegalski, M. D.; Baddorf, A. P.; Kalinin, S. V. Surface Control of Epitaxial Manganite Films via Oxygen Pressure. *ACS Nano*, **2015**, *9* (4), pp 4316–4327

(39) Fujii, S.; Ziatdinov, M.; Higashibayashi, S.; Sakurai, H.; Kiguchi, M. Bowl Inversion and Electronic Switching of Buckybowls on Gold. *Journal of the American Chemical Society* **2016**, *138*, 12142-12149.

(40) Ziatdinov, M.; Maksov, A.; Kalinin, S. V. Learning surface molecular structures via machine vision. *npj Computational Materials* **2017,** 3, 31.

(41) Kingma, D. P.; Ba, J. Adam: A Method for Stochastic Optimization. *arXiv preprint arXiv:1412.6980* **2014**.

(42) Barthel, J. (2017) *Dr. Probe - High-resolution (S)TEM image simulation software*. http://www.er-c.org/barthel/drprobe/.